\documentclass[manuscript,nonacm]{acmart}
\AtBeginDocument{%
  }

\setcopyright{none}
\usepackage{xspace}
\usepackage{multirow}
\usepackage{multicol}
\usepackage{tcolorbox}
\usepackage{graphicx}
\usepackage{subcaption}
\usepackage{listings}
\tcbuselibrary{listings}

\lstdefinelanguage{json}{
    basicstyle=\scriptsize\ttfamily, % 字体大小，根据需要调整为 \footnotesize 或 \small
    numbers=left,
    numberstyle=\tiny,
    stepnumber=1,
    numbersep=8pt,
    showstringspaces=false,
    breaklines=true,
    frame=lines,
    backgroundcolor=\color{background},
    literate=
     *{0}{{{\color{numb}0}}}{1}
      {1}{{{\color{numb}1}}}{1}
      {2}{{{\color{numb}2}}}{1}
      {3}{{{\color{numb}3}}}{1}
      {4}{{{\color{numb}4}}}{1}
      {5}{{{\color{numb}5}}}{1}
      {6}{{{\color{numb}6}}}{1}
      {7}{{{\color{numb}7}}}{1}
      {8}{{{\color{numb}8}}}{1}
      {9}{{{\color{numb}9}}}{1}
      {:}{{{\color{punct}{:}}}}{1}
      {,}{{{\color{punct}{,}}}}{1}
      {\{}{{{\color{delim}{\{}}}}{1}
      {\}}{{{\color{delim}{\}}}}}{1}
      {[}{{{\color{delim}{[}}}}{1}
      {]}{{{\color{delim}{]}}}}{1},
}

\newtcblisting{jsonbox}[2][]{
  colback=white,
  colframe=gray!20!black,
  listing only,
  listing options={
    basicstyle=\scriptsize\ttfamily,
    breaklines=true,
    language=java,
    keywordstyle=\color{blue!70!black}\bfseries,
    stringstyle=\color{red!50!black},
    commentstyle=\color{green!50!black},
    morekeywords={cluster_id, memory_unit, P, S, A, T, additional_attributes, keywords, summary},
    frame=none
  },
  title={#2},
  #1
}

\begin{document}

\title{\method: LLM-driven Formal Specification Generation with Evolving Domain Knowledge Base }

\author{Wenhan Wang}
\email{wangwenhan@isacas.ac.cn}
\affiliation{%
  \institution{Institute for Software Chinese Academy of Sciences}
  %\city{Anonymous}
  \country{China}
}
\author{Zeyu Sun}
\email{zeyu.zys@gmail.com}
\affiliation{%
  \institution{Institute for Software Chinese Academy of Sciences}
  %\city{Anonymous}
  \country{China}
}


\renewcommand{\shortauthors}{Wang et al.}

\newcommand{\method}[0]{\textsc{KBSpec}\xspace}
\begin{abstract}
Automated formal specification generation is a key step towards program understanding and formal verification. Recently, due to the success of large language models (LLMs) in code generation, researchers have made early attempts to adopt LLMs for generating formal specifications. However, the lack of formal specification language corpora in the wild often makes LLMs fail to generate syntactically correct and semantically verifiable specifications. To mitigate this gap, we propose \method, which augments LLMs with dual-source knowledge of formal specification languages: external knowledge from official documentation, and internal knowledge distilled from verifier feedback on LLM-generated specifications. \method maintains a self-evolving knowledge base that is continuously updated from successful generation and repair trajectories, without any LLM parameter tuning or labeled training data. We evaluate \method on Java Modeling Language (JML) specification generation with three LLM backends, and the results show that \method improves verification pass rates by 14–32\% over state-of-the-art LLM-based approaches, while producing the largest number of high-completeness specifications.
\end{abstract}


\maketitle

\section{Introduction}
Formal specifications serve as a key for ensuring the correctness, safety, and maintainability of software systems. They provide mathematically precise, unambiguous descriptions of system behavior and constraints instead of relying solely on informal requirements or code comments. Despite their substantial benefits, writing formal specifications remains a labor-intensive task requiring deep expertise in formal specification languages. Usually, formal specifications contain preconditions and postconditions, loop invariants and assertions, which require complex and accurate reasoning on program semantics and execution behavior. With the explosion of large language models (LLMs), researchers have leveraged the powerful reasoning abilities of LLMs in formal specification generation \cite{ma2025specgen, ma2024speceval, wen2024enchanting, le-cong-etal-2025-llms} and demonstrated performance superior to that of traditional specification generation tools \cite{ernst2007daikon, flanagan2001houdini}.

%已有方法用在这里存在什么问题？文档少是challenge(不全)，因此需要更新->引入内部知识
%While LLMs introduce promising potential for formal specification generations, several significant challenges remain. A key challenge is that LLMs still lack the domain knowledge for formal specification languages. 
However, significant challenges remain for LLM-based formal specification generation: one key challenge is the \textbf{lack of domain knowledge}.
This is mainly because formal specifications are not widely used in most software development practices, which makes them ``low-resource languages'' that scarcely exist in the training corpora of LLMs. This lack of domain knowledge causes LLMs to frequently produce low-level errors, from malformed syntax to missing boundary checks. For example, previous studies \cite{le-cong-etal-2025-llms} have found that syntax errors are the most frequent type of error in Java Modeling Language (JML) \cite{burdy2005overview} specification generation, accounting for over 40\% of the total errors in LLM-generated specifications.
%The lack of relevant knowledge makes LLMs difficult to generate correct and verifiable specifications, so the key to improve LLMs for specification generation is to integrate LLMs with more knowledge in formal specification languages.

%an empirical study found high rates of grammar/syntax errors and mis-identification of relevant program variables in LLM-generated formal DSL specifications. Second, LLMs exhibit limited semantic reasoning capabilities: they may generate plausible but incorrect assertions because they fail to reason exhaustively over all program behaviours—particularly in the presence of loops, nested control flows, or subtle invariants. Third, there is a lack of robust grounding and context-sensitivity: informal requirements, comments, and code often omit domain models, assumptions, or preconditions, and LLMs may generate overly general, ambiguous, or incomplete contracts failing to capture implied constraints or traceability. Together, these challenges — (1) syntax correctness, (2) deep semantic reasoning, and (3) domain/context grounding — limit the reliability and adoption of LLM-based specification generation in safety- or mission-critical software engineering.

To tackle the lack of domain knowledge in LLM-driven formal specification generation, a key direction is to explicitly augment LLMs with the external knowledge of formal specification languages, such as documents or tutorials written by the formal specification language developers. A similar idea has been adopted in code generation, where retrieval-augmented generation from online resources such as StackOverflow or GitHub has shown effectiveness \cite{parvez2021retrieval, yang2025empirical}. 

However, for formal specification, external knowledge resources alone are insufficient: compared to most programming languages, \textbf{the available resources for formal specification languages can be scarce or incomplete and may not cover the full diversity of program verification patterns}. For example, in the JML official tutorials provided by the OpenJML \cite{cok2011openjml} verifier developers, some topics are left unwritten as empty placeholders \footnote{https://www.openjml.org/tutorial/JavaErrorsAndExceptions}. For the ACSL specification language, the official document states that some features in the document are still experimental and unstable \footnote{https://frama-c.com/download/acsl.pdf}.

Based on these challenges, we propose \method(\textbf{\underline{K}}nowledge \textbf{\underline{B}}ase-Augmented \textbf{\underline{Spec}}ification Generation), an LLM-driven specification generation framework integrated with two knowledge sources: 1) \textbf{External knowledge} collected from officially maintained online documentation; and 2) \textbf{Internal knowledge} collected by interacting with the formal verifier using LLM-generated specifications (e.g., what types of specifications will pass verification and what types may cause certain verification errors).
%By gathering verifier feedback from a wide range of specifications, the internal knowledge acts as a complement to the external knowledge, as it can cover highly various program and specification patterns that are not included in the external knowledge resources.
External knowledge provides an initial understanding of the specification language, while internal knowledge captures verifier-compatible generation/repair patterns that are difficult to obtain from public resources alone.

%%In \method, we propose a 3-step pipeline to integrate both external and internal knowledge into a single dynamic knowledge base. First, we collect online official resources for formal specification languages, and use them to initialize a specification knowledge base. Second, we propose a knowledge base evolving strategy that iterates on a set of programs for multiple epochs: in each epoch, we perform a generation-and-repair pipeline with retrieval from the knowledge base. We collect successfull specification generation and repair records along with the verifier feedback, and use them to gradually update the knowledge base. Third, after we acquire the final knowledge base, we perform retrieval-augmented specification generation on the evaluation dataset. Our approach resembles the ``training'' process of deep learning models, but is completely parameter-free and does not require any human-annotated specification data. This makes \method agnostic to the choice of LLMs, and can be deployed to various formal specification languages without massive computational resources.

In \method, we propose a 3-step pipeline. First, we initialize a specification knowledge base from official documents and examples. Second, we evolve the knowledge base over multiple epochs by running a generation-and-repair pipeline on a training corpus, collecting successful trajectories with verifier feedback, and filtering unhelpful items. Third, we perform retrieval-augmented specification generation on the evaluation set using the learned knowledge base. Crucially, \method's knowledge base evolves entirely through verifier feedback without modifying LLM parameters, making it lightweight, model-agnostic, and deployable to any specification language with an automated verifier without relying on large-scale human-annotated datasets.

To evaluate our approach, we conduct experiments on the JML specification language with the OpenJML \cite{cok2011openjml} formal verifier. We choose the largest JML specification generation benchmark, FormalBench \cite{le-cong-etal-2025-llms}, for our experiments.
We compare \method against various baselines, including base LLMs, basic self-refine/in-context-learning, and SpecGen \cite{ma2025specgen}. 
The experiment results show that \method significantly improves specification pass rates over baseline approaches by 14.73-32.33\% and produces more high-completeness specifications than the baselines.

Our main contributions are summarized as follows:

\begin{enumerate}
    \item We propose \method, an LLM-based formal specification generation approach with a dynamically evolving knowledge base. \method can self-evolve by updating its items on a ``training'' set using only verifier feedback, without LLM parameter updates or ground-truth supervision.

    \item We identify and validate the importance of dual-source knowledge---external online documentation and verifier-grounded internal knowledge---for formal specification generation.

    \item A comprehensive evaluation demonstrates that \method can significantly improve the specification pass rate compared to the state-of-the-art LLM-driven approaches, and generate more high-completeness specifications.
\end{enumerate}

\section{Related Work}
\subsection{LLMs for Formal Specification Generation}
With the advancement of large language models, there has been a growing interest in exploring LLMs for formal specification generation. Several early works focus on specific subtypes of specification, such as postconditions and loop invariants \cite{pei2023can, endres2024can, akhond2025llm}: Pei et al. \cite{pei2023can} fine-tuned open-source LLMs to generate invariants for Java code. Endres et al. \cite{endres2024can} proposed NL2PostCond, which generates postconditions from natural language descriptions, although the generated postconditions are only Python assertions and cannot be verified by formal verifiers.

Other works aim to generate general-purpose specifications for C \cite{wen2024enchanting, yang2025integrating, chen2026beyond}, Java \cite{ma2025specgen, le-cong-etal-2025-llms}, or other programming languages.
Wen et al. \cite{wen2024enchanting} proposed AutoSpec, combining LLMs with static analysis and formal verification to generate complete C specifications using ACSL and Frama-C \cite{cuoq2012frama}, while Preguss \cite{WANG:OOPSLA2026} generates ACSL specifications with LLMs by using potential runtime error locations. Another approach, ClassInvGen \cite{sun2025classinvgen}, leverages test cases to aid LLM-driven specification generation for C++ classes. Ma et al. \cite{ma2025specgen} first introduced an LLM-based specification generation approach, SpecGen, for Java specification generation. SpecGen leverages a multi-turn conversation framework with additional mutation operators to improve the correctness of generated JML specifications. Le et al. \cite{le-cong-etal-2025-llms} proposed FormalBench, the largest benchmark to date for LLM-based JML specification generation. The authors conducted an empirical study of state-of-the-art LLMs and found that existing LLMs still have very low pass rates in generating formal specifications that are syntactically correct and verifiable.

While the above approaches focus on generating formal specifications from programs, some other works target specification generation from natural language descriptions. Req2LTL \cite{11334235} utilizes an intermediate structural language to map aerospace requirements to LTL using LLMs. Cao et al. \cite{cao2025informal} conducted a large-scale empirical study of LLMs for translating natural language requirements into formal specifications in five different formal languages. Li et al. \cite{li2025extracting} evaluated LLMs on extracting formal specifications from UAV flight control software while discovering two major limitations: specification oversimplification and specification fabrication.

\subsection{Agentic LLM Memory}
Our approach is related to the area of agentic LLM memory \cite{zhang2025survey}. These approaches augment LLMs with external memory modules that support storage, reading, and writing operations, which allow LLMs to interact with broader contexts \cite{packer2023memgpt, zhong2024memorybank, suzgun2025dynamic, xu2025amem, zhang2025agentic}. For example, Memorybank \cite{zhong2024memorybank} maintains a memory storage from historical LLM interactions.
%Similarly, Dynamic Cheatsheet \cite{suzgun2025dynamic} proposed a lightweight framework for a persistent, self-evolving LLM memory, which focuses on storing and reusing accumulated strategies and code snippets. 
MemGPT \cite{packer2023memgpt} draws inspiration from operating systems and performs memory read/write operations via LLM function calls. Xu et al. \cite{xu2025amem} proposed an agentic memory system A-Mem that creates interconnected memory ``notes'' with contextual tags and links. It can continuously refine itself as the agent performs update actions to notes and links between notes. Zhang et al. \cite{zhang2025agentic} proposed ACE (Agentic Context Engineering), which organizes and refines a memory base using 3 types of operations: generation, reflection, and curation.
%Unlike these general-purpose memory systems, \method grounds its memory updates in formal verifier feedback, ensuring that learned knowledge is validated against a rigorous correctness oracle.

\section{Motivating Example}
%1 or 2 motivating examples?
%external knowledge yes, internal knowledge?

\begin{figure*}[htbp]
  \centering
  \includegraphics[width=\linewidth]{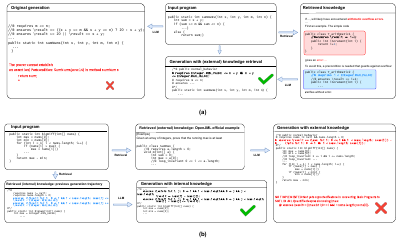}
  \caption{Motivating examples demonstrating the importance of leveraging knowledge for specification generation.}
  \label{fig:motivating}
  \Description{Examples.}
\end{figure*}

\iffalse
\begin{figure*}[t]
    \centering
    
    % 第一张子图 (a)
    \begin{subfigure}[t]{\linewidth}
        \centering
        \includegraphics[width=0.95\linewidth]{figs/kbspec-motivating-1.pdf} % 替换为你的图片文件名
        \caption{}
        \label{fig:m1}
    \end{subfigure}
    %\hfill
    % 第二张子图 (b)
    \begin{subfigure}[t]{\linewidth}
        \centering
        \includegraphics[width=0.95\linewidth]{figs/kbspec-motivating-2.pdf} % 替换为你的图片文件名
        \caption{}
        \label{fig:m2}
    \end{subfigure}
    %\hfill
    
    \caption{Motivating examples for leveraging external (a) and internal knowledge (b) for specification generation.}
    \label{fig:motivating}
\end{figure*}
\fi

Figure \ref{fig:motivating} (a) shows a motivating example for leveraging external knowledge sources. As demonstrated, \texttt{SumNums} is a function that computes the sum of variables of type \texttt{int}. If we directly ask an LLM to generate its specification, it may not know that when a program may trigger an integer overflow, the value ranges of variables must be specified in the precondition to prevent overflow. The LLM will therefore likely generate a specification that introduces a verification failure in the postcondition. In contrast, if we provide the LLM with a knowledge source from the official JML tutorial that states how to prevent the overflow bug, then the LLM is able to include the value ranges in the precondition, thus allowing the specification to be successfully verified.

Figure \ref{fig:motivating} (b) further demonstrates that external knowledge alone is not sufficient for generating correct formal specifications. Consider the Java method \texttt{bigDiff}, which computes the difference between the maximum and minimum elements of an integer array. If we only retrieve knowledge from external sources (e.g., JML official documents), we can see that the retriever returns some JML official examples related to computing max values. However, this results in a JML error where the LLM-generated specification uses the generalized quantifiers \texttt{\textbackslash max} and \texttt{\textbackslash min}. This is because the OpenJML verifier in ESC mode does not fully support some generalized quantifiers \cite{le-cong-etal-2025-llms}, but this limitation is not fully documented or explained, nor are alternative encodings provided. By contrast, with our approach, which can retrieve internal knowledge gained from past successful generation and repair patterns, the LLM learns how to avoid the \texttt{\textbackslash max} quantifier by generating a specification using \texttt{\textbackslash forall}, which can successfully pass OpenJML verification.

These examples together demonstrate that in formal specification generation, external knowledge resources and internal knowledge learned from past generation/repair trajectories are both important, which further motivates our approach of the self-evolving knowledge base.

\section{Approach}

\begin{figure*}[h]
  \centering
  \includegraphics[width=0.85\linewidth]{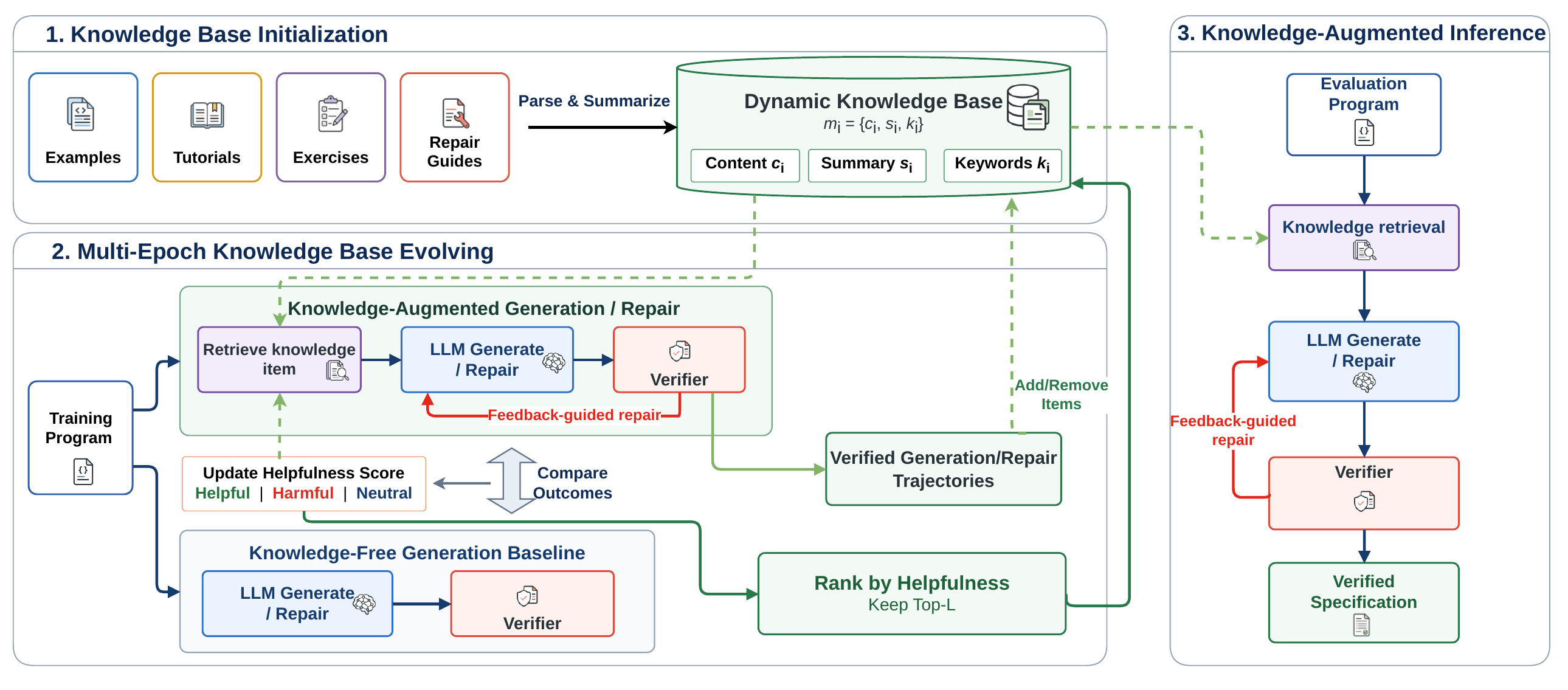}
  \caption{Overview of \method.}
  \label{fig:overview}
  \Description{Overview.}
\end{figure*}

Figure \ref{fig:overview} shows the overall structure of our approach. \method is an LLM-based framework that involves interacting with a dynamically updated knowledge base $M$ and a formal verifier. The whole pipeline consists of three steps: 

\begin{enumerate}
    \item \textbf{Knowledge base initialization}: In this step, we build an initial knowledge base for formal specification using external knowledge sources, such as official documents and example specifications. This knowledge base will further support retrieval-augmented specification generation and repair.

    \item \textbf{Knowledge base evolution with verifier feedback}: In this stage, we propose a self-evolving mechanism to update the knowledge base by interacting with the formal verifier. Given a ``training'' corpus with Java programs, \method performs a generation-repair pipeline based on the verifier feedback, and continuously updates and filters the knowledge base using the trajectories of successful generation or error fixing patterns. Notice that our knowledge base evolution process does not change the parameters of the LLM itself, thus it can be applied to most LLMs without high adaptation costs or massive computational resources.

    \item \textbf{Knowledge-augmented inference}: In this step, we perform specification generation and repair on the evaluation dataset with the aid of the learned knowledge base.
\end{enumerate}

\subsection{Specification Knowledge Base: Initialization}

\begin{table}[t]
\centering
\begin{tabular}{lc}
\toprule
    Data Source & No. items \\
\midrule
    OpenJML examples & 11 \\
    OpenJML tutorials & 31 \\
    OpenJML exercises & 26 \\
    Repair guidelines \cite{le-cong-etal-2025-llms} & 15 \\
\midrule
    Total & 83 \\
\bottomrule
\end{tabular}
\caption{Distribution of the data sources for building the initial knowledge base.}
\label{tab:knowledge-source}
\end{table}

We utilize the official OpenJML website \footnote{https://www.openjml.org/} to gather the initial knowledge of JML. This website includes introductions to the JML language and the OpenJML verifier, as well as the official tutorial on writing JML specifications. Specifically, we collect the tutorial documents under the ``tutorial'' column and the expert-written JML specification examples under the ``examples'' column.
%In total, we collect 31 tutorial pages and 11 official examples. 
Then we reformat these web pages into MarkDown files and remove the content not related to JML specifications. These examples and tutorials are further used to initialize the knowledge base of {\method}. For all official examples, we follow FormalBench \cite{le-cong-etal-2025-llms} and rewrite them with both the code before and after specification generation. We further include the OpenJML supplementary exercises: each exercise contains one or two questions related to writing certain types of specifications or analyzing existing specifications with certain error patterns. For these exercises, we split each document item into separate questions. We also add the bug-fixing guidelines provided by FormalBench, which consists of human-written guidelines for different types of JML errors. The statistics for all external knowledge sources are demonstrated in Table \ref{tab:knowledge-source}

Each knowledge item $m_{i}$ in the \method knowledge base $M$ is represented as:

\begin{equation}
    m_{i} = \{c_{i}, s_{i}, k_{i}\}
\end{equation}

Here, $c_{i}$ is the original knowledge content (e.g., example or document), $s_{i}$ is a shorter summary of the knowledge content, and $k_{i}$ is a set of keywords containing the core JML concepts in $c_{i}$. Both $s_{i}$ and $k_{i}$ are generated using an LLM from $c_{i}$. A similar hierarchical knowledge representation has been adopted in work on agentic memory \cite{xu2025amem}. These three fields represent different abstraction levels for knowledge in generating or repairing formal specifications. By summarizing the original knowledge content, we can provide the LLM with high-level domain knowledge and concepts for generating correct specifications.

\iffalse
\begin{figure}[h]
  \centering
  \includegraphics[width=\linewidth]{figs/training-group.pdf}
  \caption{The training strategy for tagging retrieved knowledge items.}
  \label{fig:train}
  \Description{Update mechanism.}
\end{figure}
\fi

\subsection{Knowledge Base Evolving with Verifier Feedback}

In this stage, we perform a retrieval-augmented specification generation and repair pipeline which runs for multiple epochs. The key insight is to update the knowledge base by selected specification generation/repair trajectories, and remove knowledge items that are not helpful to generate correct specifications. The detailed pipeline in one \method epoch consists of the following steps:

\subsubsection{Epoch Initialization}
At the beginning of each epoch, we first initialize a candidate new knowledge set $C_{epoch}$ that stores new knowledge items generated during this epoch.
For the knowledge items already in the knowledge base, in order to determine if an item is helpful, harmful, or neutral, each knowledge item $m_{i}$ maintains a helpfulness score triplet $<s_{helpful}, s_{harmful}, s_{neutral}>$. At the beginning of each knowledge evolving epoch, all knowledge items' helpfulness scores are initialized as $<s_{helpful}=0, s_{harmful}=0, s_{neutral}=0>$.

\subsubsection{Initial Specification Generation}

For a given program $P$ to be verified, we first generate a baseline specification $S_{base}=\textrm{LLM}(P)$ with the plain LLM following the generate-and-repair pipeline of \cite{le-cong-etal-2025-llms}, without any retrieved knowledge. $S_{base}$ will further help us in determining whether a retrieved knowledge item is helpful or not. Then we use $P$ to query the knowledge base. We use a BERT-based embedding model $\textrm{embed}()$ as the retriever, and the retrieval similarity score for $P$ and the knowledge item $m_{i}$ is computed by:

\begin{equation}
     sim(P,m_{i})=\textrm{cos}(\textrm{embed}(P),\textrm{embed}(c_{i} \parallel s_{i} \parallel k_{i}))
\end{equation}

Here, the original knowledge content, the summary, and the keywords are all concatenated to compute the embedding. The retriever selects the $G$ knowledge items with the highest similarity scores, denoted as $m_1, \ldots, m_G$. We then generate $G$ different specifications $S_1, \ldots, S_G$, each using one selected knowledge item as an additional source: $S_j=\textrm{LLM}(P,m_j)$. This design makes each verification outcome attributable to a unique knowledge item.

After we generate all specifications for $P$, we verify them and update the helpfulness score based on verification results. Let $\textrm{Verify}(S) = \textrm{True}$ denote that specification $S$ passes verification, and $\textrm{Verify}(S) = \textrm{False}$ denote that $S$ fails verification. For each generated specification $S$ such that $\textrm{Verify}(S) = \textrm{True}$, we add its generation trajectory $<P \to S>$ to the candidate new knowledge set $C_{epoch}$. The helpfulness score of each selected knowledge item $m_j$ is then updated by the following transition function $\tau(m_j)$:

\begin{equation}
\resizebox{0.9\linewidth}{!}{$
%\label{eq:sigma_definition}
    \tau(m_{j}) = 
    \begin{cases} 
        s_{helpful} += 1 & \text{if } \textrm{Verify}(S_{j}) = \textrm{True}  \And  \textrm{Verify}(S_{base}) = \textrm{False}, \\
        s_{harmful} += 1 & \text{if } \textrm{Verify}(S_{j}) = \textrm{False}  \And  \textrm{Verify}(S_{base}) = \textrm{True}, \\
        s_{neutral} += 1 & \text{otherwise},
    \end{cases}
$}
\end{equation}

The design of the helpfulness score is based on the idea that if an LLM cannot generate a correct specification for $P$ without extra knowledge, while it generates a correct specification after receiving additional knowledge, then the knowledge item is considered helpful, and vice versa.

\subsubsection{Iterative Repair}

After the initial specification generation step, \method has gathered some knowledge from the successful one-time generation trajectories, but more experience can be learned by repairing the incorrect specifications. For each initially failed specification $S_j$, we set $S_j^{(0)}=S_j$ and perform at most three repair rounds $t\in\{0,1,2\}$. In round $t$, we query the knowledge base with the current specification $S_j^{(t)}$ and its verifier error message, and use the single most relevant item $m_j^{(t)}$ for retrieval-guided repair. The next specification is $S_j^{(t+1)}=\textrm{LLM}(S_j^{(t)},\textrm{Err}(S_j^{(t)}),m_j^{(t)})$. The process terminates when $S_j^{(t+1)}$ verifies or after three unsuccessful rounds. The single-item design attributes each repair outcome to one knowledge item.

Sometimes, a buggy specification is not fixed by a single repair round, but with multiple continuous fixing steps, each step fixes a part of errors. To measure partial repair progress, let $E_t$ denote the error-type multiset extracted from $\textrm{Err}(S_j^{(t)})$. Repeated occurrences of the same normalized error type remain separate multiset elements. We define the resolved-obligation ratio as:

\begin{equation}
    p_t = \frac{|E_t-E_{t+1}|}{|E_t|}
        = \frac{\sum_e \max(0,\operatorname{count}_{E_t}(e)-\operatorname{count}_{E_{t+1}}(e))}{|E_t|}.
\end{equation}

This ratio rewards only error instances that disappear after the repair. Newly reported error types are not immediately treated as harmful because resolving an earlier failure may allow OpenJML to expose deeper proof obligations. We initialize the multiset of previously resolved error instances as $H_0=\emptyset$. After an unsuccessful round with usable error multisets, we update it as $H_{t+1}=H_t\uplus(E_t-E_{t+1})$, where $\uplus$ denotes multiset union; otherwise, $H_{t+1}=H_t$. If a previously resolved error reappears, its regression penalty is:

\begin{equation}
    q_t = \frac{|H_t \cap E_{t+1}|}{\max(|H_t|,1)}.
\end{equation}

For the same program, we reuse the precomputed outcome of the selected baseline, which starts from its own generated specification and performs at most three verifier-feedback repair rounds without retrieved knowledge-base items. Let $B_{fix}$ be true iff this baseline produces a verified specification during these repair rounds. For the single retrieved item $m_j^{(t)}$, we update the score triplet using the following baseline-aware rule:

\begin{equation}
\resizebox{0.96\linewidth}{!}{$
    \tau(m_j^{(t)}) =
    \begin{cases}
        s_{helpful} += 1
            & \text{if } \textrm{Verify}(S_j^{(t+1)})=\textrm{True} \And B_{fix}=\textrm{False}, \\
        s_{neutral} += 1
            & \text{if } \textrm{Verify}(S_j^{(t+1)})=\textrm{True} \And B_{fix}=\textrm{True}, \\
        s_{harmful} += 1
            & \text{if } \textrm{Verify}(S_j^{(t+1)})=\textrm{False} \And B_{fix}=\textrm{True}, \\
        (s_{helpful},s_{harmful},s_{neutral}) += (p_t,q_t,1-p_t)
            & \text{if both repairs fail and } |E_t|>0, \\
        s_{neutral} += 1
            & \text{if both repairs fail without a usable error multiset}.
    \end{cases}
$}
\end{equation}

Timeouts, internal errors, and verifier outputs without extractable error types do not create artificial obligations: they are neutral when the baseline repair also fails and harmful when the baseline succeeds. A successful repair receives only its terminal credit and does not additionally accumulate partial credit. If $S_j^{(t+1)}$ successfully verifies without errors, we add its successful repair trajectory to the candidate knowledge set $C_{epoch}$.

\subsubsection{Knowledge Base Update}

In KBSpec, we use a multi-epoch knowledge base strategy: at the end of each ``training'' epoch, after we run the generation and repair pipeline on all programs in the ``training'' set, we update the knowledge base with the gathered helpfulness scores and the new candidate set $C_{epoch}$.

First, for the items already in the knowledge base, we set a knowledge evolving limit $L$ to filter down the knowledge base. We keep the $L$ items with the highest helpfulness score $r_{helpful}$:

\begin{equation}
     r_{helpful} = \frac{s_{helpful} - s_{harmful}}{s_{helpful}+s_{harmful}+s_{neutral}+1e-6}
\end{equation}

The score normalizes the difference between helpful and harmful feedback by the total amount of feedback; neutral feedback reduces the magnitude of the score. The $1e-6$ is added to prevent division by zero. Then, we add the items in $C_{epoch}$ to the knowledge base. To prevent the redundancy of knowledge items, we perform a similarity check: before adding a new knowledge item $m$, we first use its content to query the knowledge base using the retriever. If there exists an item $\hat{m}$ such that $sim(m, \hat{m})>0.95$, we consider this new item redundant and do not add it to the knowledge base. Otherwise, we generate the summary and keywords for this item, and add it to the knowledge base.

 We continuously run the knowledge update on the ``training'' set for a number of epochs. After each epoch of training, we evaluate the final specification pass rate (after fixing) on the epoch's training set. Similar to previous experiences in deep learning training, we terminate the ``training'' early if the pass rate does not increase in 2 epochs to avoid ``overfitting''.

%TODO: 添加prompt template示例？有空间的话

\subsection{Knowledge-Augmented Inference}
In the inference stage of \method, we apply a similar retrieval-augmented generation and repair pipeline without updating the knowledge scores. For initial generation, we use the source program as the query and directly select the top 3 knowledge items ranked by the dense retriever. For each repair round, we query the same dense retriever with the current specification and its OpenJML error message, and again use the top 3 ranked items. If fewer than 3 items are available, we use all returned items. Similar to the knowledge evolving stage, if the initial specification generation is unsuccessful, we perform iterative repair for at most 3 rounds.

\section{Evaluation}

\subsection{Research Questions}

To evaluate and understand {\method} for JML specification generation, we aim to answer the following research questions:

\begin{itemize}
    \item \textbf{RQ1:} How does {\method} improve the effectiveness of specification generation over state-of-the-art baselines?

    %\item \textbf{RQ2:} What are the contributions of leveraging external knowledge sources?

    %\item \textbf{RQ3:} How does the self-evolving knowledge base helpful to specification generation?

    %\item \textbf{RQ4:} What kinds of knowledge are helpful in generating correct specifications?

    \item \textbf{RQ2:} How does each component of \method affect the performance of specification generation?

    \item \textbf{RQ3:} What kinds of knowledge are acquired during knowledge base evolution, and which of them are the most useful?

    \item \textbf{RQ4:} How well does the knowledge base learned by \method generalize to a different specification generation benchmark?
\end{itemize}

\subsection{Experiment Settings}
\subsubsection{Dataset}
We conduct our main experiments on the FormalBench \cite{le-cong-etal-2025-llms} dataset. FormalBench is a state-of-the-art benchmark for JML specification generation from Java source code. The dataset consists of two parts: the FormalBench-Base dataset, which contains 699 Java programs for evaluation, and the FormalBench-Diverse dataset, which comprises 6,219 programs that are built from applying mutations on FormalBench-Base. In this paper, we follow previous settings and use the FormalBench-Base dataset for evaluation. We use a subset of FormalBench-Diverse (containing 1,794 programs with complete metadata, following the settings of the FormalBench repository) only for knowledge base evolution prior to evaluation. In each epoch, we only randomly sample \textbf{100} programs from the diverse dataset for training, and the training stage continues for up to 10 epochs under the early-stopping mechanism.

\subsubsection{Implementation Details}
We select three different LLMs as the base model for \method: GPT-5.2 \cite{openai2025gpt52}, GPT-5-mini \cite{openai2025gpt5}, and Claude-Sonnet-5.
In the knowledge retrieval stage, we adopt a BERT-based embedding model, all-minilm-l6-v2 provided by sentence-transformers \cite{reimers2019sentence}, as the dense retriever. We use OpenJML 21 as the formal verifier, which is the same as FormalBench. 
During training, the retriever selects the $G=5$ knowledge items with the highest relevance scores for separate rollouts.

\subsubsection{Evaluation Metrics}
We follow existing works \cite{le-cong-etal-2025-llms, ma2025specgen} on LLM-based JML specification generation and evaluate the quality of generated specifications using the following metrics:

\begin{itemize}
    \item Pass rate (PR): the rate of generated specifications that can be successfully verified by the verifier.

    \item Fail rate (FR): the rate of generated specifications that trigger at least one error by the verifier. Failed verification can be caused by many reasons, such as syntax errors or incorrect invariants. Notice that in most circumstances $\textrm{PR}+\textrm{FR} < 1$, because the verifier may return other statuses besides verification failure, such as timeout or internal JML errors (which may be caused by bugs or unimplemented features of OpenJML).

    \item Completeness@Pass (C@Pass): a high-quality specification should be both correct and \textbf{complete}, i.e., it should accurately specify all possible behavior of the program. We adopt the mutation-based completeness metric from \cite{le-cong-etal-2025-llms}: first apply mutation testing to the program to generate $K$ non-equivalent mutants, then verify the generated specification on these mutants. Theoretically, if the original specification is complete, all mutants should \textbf{fail} verification. Therefore, if $n_{failed}$ mutants do not pass, the completeness score is computed by $\frac{n_{failed}}{K}$. C@Pass averages this score over the specifications that pass verification. The mutants are generated with Major mutator using predefined mutation rules \cite{le-cong-etal-2025-llms}. It further uses the state-of-the-art Equivalent Mutant Suppression (EMS) technique to filter equivalent mutants; its manual inspection found only 3 of 232 mutants (1.3\%) to be equivalent, indicating that the retained mutants are overwhelmingly non-equivalent \cite{le-cong-etal-2025-llms}.

    \item Completeness@All (C@All): C@Pass alone cannot fully reflect specification generation capability because it considers only programs for which a method succeeds: a method may obtain a high C@Pass by generating verifiable specifications for only a small subset of easy programs. For specification generation, producing a verifiable specification with lower completeness is still more meaningful than producing no verifiable specification. We therefore introduce C@All to balance pass rate and completeness over the entire dataset by assigning a completeness score of zero to every program without a verifiable specification. For a dataset of $N$ programs, C@All is defined as $\mathrm{C@All}=\frac{1}{N}\sum_{i=1}^{N}c_i$, where $c_i$ is the mutation-based completeness of the final specification for program $i$. 
\end{itemize}

\subsubsection{Baselines}
We first evaluate our approach with three different LLM backends: GPT-5.2, GPT-5-mini, and Claude-Sonnet-5. In this setting, we adopt the same basic prompt template as FormalBench \cite{le-cong-etal-2025-llms}. We also choose SpecGen \cite{ma2025specgen}, the state-of-the-art JML specification generation approach, as an additional baseline. SpecGen generates a JML specification by performing a multi-turn conversation with an LLM and several mutation operations if the LLM-generated specification fails to verify. Furthermore, we compare \method against several advanced LLM prompting techniques examined by \cite{le-cong-etal-2025-llms}:

\begin{itemize}
    \item \textbf{Few-shot in-context-learning}: The LLM is prompted by two expert-written examples on how to generate JML specifications from Java code.

    \item \textbf{Least-to-Most (LTM) prompting}: LTM \cite{zhou2023leasttomost} is an advanced CoT prompting technique that breaks down a complex task into smaller subtasks. For this approach, we use the same prompt template as \cite{le-cong-etal-2025-llms}: it splits the JML specification generation task into 3 steps: 1) generate the weakest precondition; 2) generate the strongest postcondition; and 3) generate the rest of the specifications, e.g., loop invariants and assertions. The prompt template further provides two examples of how to generate JML specifications step by step with natural language explanations.

    \item \textbf{Self-repair with error type analysis (``+Repair'' in Table \ref{tab:results})}: FormalBench further introduced a self-repair mechanism based on the verifier error message: the authors first extract error type information from the error message and provide corresponding guidance for repairing certain types of errors. The self-repair can be applied to both few-shot and LTM specification generation approaches.
\end{itemize}

%\subsection{Experiment Results}

\subsection{RQ1: Overall Performance}

\begin{table*}[t]
\centering
\resizebox{\textwidth}{!}{%
\begin{tabular}{lcccc c cccc c cccc}
\toprule
\multirow{2}{*}{Approach}
  & \multicolumn{4}{c}{GPT-5.2}
  & 
  & \multicolumn{4}{c}{GPT-5-mini}
  &
  & \multicolumn{4}{c}{Claude-Sonnet-5} \\
\cline{2-5}\cline{7-10}\cline{12-15}
  & PR & FR & C@Pass & C@All
  &
  & PR & FR & C@Pass & C@All
  &
  & PR & FR & C@Pass & C@All \\
\cline{2-5}\cline{7-10}\cline{12-15}
Base     & 10.73 & 58.37 & 85.73 & 9.20  &  & 5.59 & 65.47 & 89.11 & 4.98  &  & 5.72 & 80.11 & \textbf{94.28} & 5.39 \\
Few-shot & 14.88 & 51.07 & 87.65 & 13.04 &  & 7.58 & 64.94 & \textbf{92.55} & 7.02  &  & 20.89 & 46.78 & 79.23 & 16.55 \\
LTM      & 14.35 & 55.09 & \textbf{90.94} & 13.05 &  & 9.58 & 60.80 & 88.94 & 8.52  &  & 19.31 & 45.92 & 84.09 & 16.24 \\
+Repair  & 30.90 & 37.05 & 78.92 & 24.39 &  & 18.31 & 71.24 & 81.54 & 14.93 &  & 27.18 & 22.60 & 80.92 & 22.00 \\
\midrule
SpecGen  & 41.63 & 32.90 & 64.04 & 26.66 &  & 28.18 & 36.05 & 74.11 & 20.88 &  & 47.93 & 27.32 & 64.16 & 30.75 \\
\midrule
\method  & \textbf{73.96} & \textbf{13.45} & 52.06 & \textbf{38.51} &  & \textbf{47.35} & \textbf{28.61} & 65.30 & \textbf{30.92} &  & \textbf{62.66} & \textbf{16.31} & 62.97 & \textbf{39.46} \\
\bottomrule
\end{tabular}
}
\caption{Results of \method compared to baseline approaches with different LLM backends. All metrics are reported in \%.}
\label{tab:results}
\end{table*}

Table \ref{tab:results} presents the overall performance of \method over baseline approaches on FormalBench-Base across three LLM backends (GPT-5.2, GPT-5-mini, and Claude-Sonnet-5). In general, \method consistently achieves the highest verification pass rate, the lowest fail rate, and the highest C@All across all models, substantially outperforming both advanced prompting baselines and the state-of-the-art baseline SpecGen.

On GPT-5.2, \method achieves a PR of 73.96\%, improving over SpecGen (41.63\%) by 32.33 percentage points and over the base setting by 63.23 percentage points. Similar trends are observed for GPT-5-mini and Claude-Sonnet-5, where \method improves the PR over SpecGen by 19.17 and 14.73 percentage points, respectively. Similar trends can be found in the failure rates: \method substantially reduces the failure rate across all three models. For GPT-5.2, \method successfully reduces the fail rate to a low value of 13.45\%.
These improvements demonstrate that \method generalizes across both flagship and cost-efficient LLMs. 

Note that the sum of PR and FR is not 1, which means that the generated specification may also encounter other types of errors, most of which are timeout errors. The results demonstrate that \method is also capable of reducing timeout errors along with verification failures. 

%%In addition to increasing pass rates, MemSpec reduces verifier failures. For example, on GPT-5.2, FR decreases from 31.33\% (SpecGen) to 22.75\% (MemSpec); on GPT-5-mini, FR drops sharply from 46.78\% to 19.03\%. This indicates that MemSpec not only increases successful verification outcomes but also meaningfully mitigates common sources of verification failure, such as syntax errors and inconsistent specifications. Compared to advanced prompting strategies (Few-shot, LTM, and +repair), MemSpec yields consistently higher PR across all models, suggesting that prompt engineering and multi-turn repair alone are insufficient for addressing the knowledge gap inherent in low-resource formal specification languages.

Regarding specification strength, C@Pass results reveal a nuanced picture. SpecGen retains a higher C@Pass than \method on all three backends, with differences of 11.98, 8.81, and 1.19 percentage points on GPT-5.2, GPT-5-mini, and Claude-Sonnet-5, respectively. However, \method achieves a higher C@All on all three backends, improving over SpecGen by 11.85, 10.04, and 8.71 percentage points. This shows that \method significantly broadens verification coverage while preserving a substantial absolute amount of specification completeness. We further analyze this tradeoff below, where the completeness distribution shows that \method consistently generates more high-completeness specifications.
%Nevertheless, even in this case, the substantial gain in pass rate suggests that \method primarily addresses correctness bottlenecks rather than merely relaxing specifications indiscriminately.

Taken together, the results provide strong evidence that integrating documentation-derived knowledge with verifier-feedback-driven memory evolution substantially enhances LLM-based JML specification generation. \method consistently improves verification pass rates and C@All over both raw LLM baselines and prior state-of-the-art approaches, while retaining a substantial C@Pass. These findings affirm the effectiveness and generality of the proposed knowledge-augmented framework.
%init llm version, need improving

\begin{table}[htbp]
\centering
\scalebox{0.8}{
\begin{tabular}{llccccc}
    \toprule
    % 第一行表头：Model 和 Approach 纵向合并2行，Completeness 横向合并4列
    \multirow{2}{*}{Model} & \multirow{2}{*}{Approach} & \multicolumn{5}{c}{Completeness@pass} \\
    % 在第3到第6列下方画一条较细的横线
    \cmidrule(lr){3-7}
    % 第二行表头：前两列留空（因为已经被上一行的 multirow 占据）
    & & [0, 0.25) & [0.25, 0.5) & [0.5, 0.75) & [0.75, 1] & [0.5, 1] \\
    \midrule
    % 第一组数据：GPT-5.2，纵向合并6行
    \multirow{6}{*}{GPT-5.2} 
    & Base     & 2 & 5 & 8 & 60 & 68\\
    & Few-shot & 3 & 4 & 9 & 88 & 97\\
    & LTM      & 1 & 2 & 10 & 87 & 97\\
    & Repair   & 12 & 29 & 23 & 152 & 175\\
    & SpecGen  & 50 & 35 & 49 & 157 & 206\\
    & \method  & 148 & 100 & \textbf{92} & \textbf{177} & \textbf{269}\\
    \midrule
    % 第二组数据：GPT-5-mini，纵向合并6行
    \multirow{6}{*}{GPT-5-mini} 
    & Base     & 2 & 0 & 3 & 34 & 37\\
    & Few-shot & 1 & 0 & 4 & 48 & 52\\
    & LTM      & 1 & 2 & 8 & 56 & 66\\
    & Repair   & 5 & 13 & 14 & 96 & 110\\
    & SpecGen  & 23 & 16 & 26 & 132 & 158\\
    & \method  & 39 & 70 & \textbf{57} & \textbf{165} & \textbf{222}\\
    \midrule
    % 第三组数据：Claude-Sonnet-5，纵向合并6行
    \multirow{6}{*}{Claude-Sonnet-5}
    & Base     & 0 & 0 & 6 & 34 & 40\\
    & Few-shot & 11 & 11 & 25 & 99 & 124\\
    & LTM      & 5 & 9 & 20 & 101 & 121\\
    & Repair   & 11 & 9 & 36 & 134 & 170\\
    & SpecGen  & 55 & 42 & 52 & 186 & 238\\
    & \method  & 54 & 94 & \textbf{98} & \textbf{192} & \textbf{290}\\
    \bottomrule
\end{tabular}
}
\caption{The completeness distribution of all passed specifications generated by \method and baselines.}
\label{tab:completeness-detail}
\end{table}

\subsubsection{Analysis on Completeness}
Table \ref{tab:completeness-detail} further clarifies the completeness results by showing the distribution over all passed specifications. Although \method introduces more low-completeness cases, it consistently yields the largest number of medium-to-high-quality specifications across all backbones. In particular, the number of passed specifications with completeness in [0.5, 1] rises from 206 to 269 on GPT-5.2, from 158 to 222 on GPT-5-mini, and from 238 to 290 on Claude-Sonnet-5 when compared with SpecGen. The same trend holds for highly complete specifications in [0.75, 1], where \method obtains 177, 165, and 192 cases, compared with 157, 132, and 186 for SpecGen. These results suggest that \method does not simply improve pass rate by generating trivially weak contracts; rather, it converts more previously failing cases into verifiable specifications, while still preserving a large number of semantically useful ones. This observation is consistent with Table \ref{tab:results}: the lower C@Pass is mainly a consequence of broader coverage among passed cases, while the higher C@All reflects the improved completeness yield over the entire dataset.

\begin{tcolorbox}[colback=gray!10, colframe=gray!50, boxrule=0.5pt, arc=2pt, boxsep=0pt,]
\textbf{Answer to RQ1:} \method achieves the highest pass rate, lowest fail rate, and highest C@All across all LLM backends, significantly outperforming existing LLM baseline approaches. Although C@Pass is lower than SpecGen, further analysis shows that \method produces the largest number of high completeness specifications, indicating that the gains in PR stem from broader coverage of previously failing cases rather than generating trivially weak specifications.
\end{tcolorbox}

\subsection{RQ2: Ablation Study}

RQ2 evaluates the contribution of five \method components: 1) knowledge base initialization with documents, 2) knowledge base evolution, 3) credit-based filtering for selecting helpful knowledge items, 4) partial repair credit during knowledge base evolution, and 5) iterative repair during inference. We conduct an ablation study on the above components with all three LLM backends, and the results are shown in Table \ref{tab:ablation}. The w/o partial repair credit variant retains only the terminal outcome credit and removes the partial repair credit. From the results, we find that:

\begin{table*}[t]
\centering
\resizebox{\textwidth}{!}{%
\begin{tabular}{l cccc c cccc c cccc}
\toprule
\multirow{2}{*}{Approach} & \multicolumn{4}{c}{GPT-5.2} && \multicolumn{4}{c}{GPT-5-mini} && \multicolumn{4}{c}{Claude-Sonnet-5} \\
\cline{2-5} \cline{7-10} \cline{12-15}
& PR & FR & C@Pass & C@All && PR & FR & C@Pass & C@All && PR & FR & C@Pass & C@All \\
\midrule
w/o initialization       & 74.11 & 12.45 & 48.60 & 36.01 && 45.92 & 29.18 & 64.54 & 29.64 && 53.65 & 23.03 & 66.23 & 35.53 \\
w/o evolving             & 56.94 & 27.18 & 58.51 & 33.32 && 17.17 & 36.34 & \textbf{78.87} & 13.54 && 27.75 & 42.20 & \textbf{81.98} & 22.75 \\
w/o credit-based filtering & \textbf{74.54} & \textbf{11.59} & 46.91 & 34.96 && \textbf{52.50} & \textbf{25.04} & 57.37 & 30.12 && 61.09 & 20.74 & 62.94 & 38.45 \\
w/o partial repair credit & 70.67 & 13.88 & 52.33 & 36.98 && 38.05 & 40.34 & 74.77 & 28.45 && 55.51 & 19.60 & 64.96 & 36.06 \\
w/o iterative repair     & 21.89 & 39.34 & \textbf{76.24} & 16.69 && 18.60 & 50.64 & 73.66 & 13.70 && 39.20 & 29.90 & 66.66 & 26.13 \\
\midrule
full \method    & 73.96 & 13.45 & 52.06 & \textbf{38.51} && 47.35 & 28.61 & 65.30 & \textbf{30.92} && \textbf{62.66} & \textbf{16.31} & 62.97 & \textbf{39.46} \\
\bottomrule
\end{tabular}
}
\caption{Results of the ablation study with different LLM backends.}
\label{tab:ablation}
\end{table*}

\textbf{Knowledge base evolving and iterative repair are the most impactful components.}
Removing multi-turn repair during inference causes the largest degradation in C@All on GPT-5.2 and one of the two largest degradations on the other models, with drops of 21.82, 17.22, and 13.33 percentage points on GPT-5.2, GPT-5-mini, and Claude-Sonnet-5, respectively. Removing verifier-feedback-driven evolving causes the largest degradation on GPT-5-mini and Claude-Sonnet-5, reducing C@All by 17.38 and 16.71 percentage points. This confirms that experiential knowledge distilled from generation and repair trajectories is a primary driver of \method's effectiveness, and that external documentation alone is insufficient. The iterative-repair results further indicate that a substantial portion of successfully verified specifications are obtained through knowledge-guided repair rather than first-attempt generation.

\textbf{Document-based initialization and partial credit provide a solid foundation.}
Removing document-based initialization reduces C@All by 1.28--3.93 percentage points. Removing partial repair credit also reduces C@All by 1.53--3.40 percentage points, showing that partial repair credit contributes useful learning signals before a repair reaches terminal success. These effects are smaller than those of knowledge evolving and iterative repair, but remain consistent across all three backends.

\textbf{Credit-based knowledge base filtering offers complementary benefits.}
Removing credit-based filtering reduces C@All by 0.80--3.55 percentage points, although it can slightly increase PR on GPT-5.2 and GPT-5-mini. This shows why PR and C@Pass should be considered jointly through C@All. Moreover, the goal of knowledge filtering is not only to improve specification quality, but also to restrict the size of the knowledge base and maintain retrieval efficiency: without filtering, the final knowledge bases contain 1,385, 1,669, and 814 items for GPT-5.2, GPT-5-mini, and Claude-Sonnet-5, respectively, compared with the size limit of 300 for the full method.

% claude %In summary, the training mechanism and iterative repair are the two most critical components, each contributing roughly 20--30 points of pass rate improvement. Document initialization and LLM-based retrieval provide complementary benefits whose magnitude varies with model capability, while knowledge filtering offers a modest but consistent contribution.

\begin{tcolorbox}[colback=gray!10, colframe=gray!50, boxrule=0.5pt, arc=2pt, boxsep=0pt,]
\textbf{Answer to RQ2:} Each component in \method contributes to completeness in C@All across all LLM backends, and most of them contributes to pass rates. Iterative repair and knowledge base evolution have the largest effects, while partial repair credit, document initialization, and credit-based filtering provide smaller but consistent benefits.
\end{tcolorbox}

\subsection{RQ3: Analysis on Knowledge Learning}

To analyze the knowledge learned during the knowledge base evolution process, we aim to investigate:

\begin{itemize}

    \item What types of knowledge items exist in the final knowledge base, and which types are retrieved? %investigating retrieval: both training and inference

    %\item What types of knowledge are helpful, while what types are harmful?

    \item What kinds of knowledge patterns (e.g., specification bug and repair patterns) are learned?
\end{itemize}

\begin{table*}[t]
\centering
% 如果因为列数过多导致表格超出页面宽度，可以取消下方 \resizebox 的注释
\resizebox{\textwidth}{!}{
\begin{tabular}{lccccccc}
\toprule
    \multirow{2}{*}{Model} & \multicolumn{4}{c}{External} & \multicolumn{2}{c}{Internal} \\
    \cmidrule(lr){2-5} \cmidrule(lr){6-7} % 在主分类下添加细横线区分
    & OpenJML examples & OpenJML tutorials & OpenJML exercises & Repair guidelines \cite{le-cong-etal-2025-llms} &  Generation & Repair \\
\midrule
    GPT-5.2       & 2.4 & 8.1 & 6.1 & 3.7 & 21.2 & 58.5 \\
    GPT-5-mini    & 0.2 & 0.4 & 0.6 & 0.4 & 13.7 & 84.7 \\
    Claude-Sonnet-5 & 2.2 & 8.1 & 5.9 & 3.7 & 31.5 & 48.8 \\
\bottomrule
\end{tabular}
}
\caption{Distribution of the data sources for the final learned knowledge base of \method. Results are reported in percentage proportions.}
\label{tab:mem-source}
\end{table*}

\subsubsection{Analysis on Knowledge Stored and Retrieved}

Table~\ref{tab:mem-source} reports the composition of the final learned knowledge base after training completes, revealing what types of knowledge items are gathered through the training process. Across all three backends, internally learned items overwhelmingly dominate the knowledge base. GPT-5.2 and Claude-Sonnet-5 retain similar proportions of external knowledge, while GPT-5-mini prunes most external knowledge during training: external items account for only 1.65\% of its final knowledge base.

Among the two types of internal knowledge, repair-derived items consistently outnumber generation-derived items on all three backends, confirming that successful specification repair is more frequent than successful one-time generation during training. The dominance of repair-derived knowledge is most pronounced for GPT-5-mini (84.68\% vs.\ 13.66\%).
%From these analysis results, we find that internal knowledge from verifier feedback is the main source of knowledge storage and retrieval, while external knowledge from online resources mainly serves as a seed to empower the knowledge learning mechanism.

%claude %producing a larger volume of repair attempts but with lower individual success rates, thereby limiting the accumulation of high-quality repair items. 

\iffalse
%图片适当缩小，字体调大
\begin{figure}[htbp]
    \centering
    
    % 第一张子图 (a)
    \begin{subfigure}[b]{0.4\textwidth}
        \centering
        \includegraphics[width=0.85\textwidth]{figs/gpt_5_2_retrieved_sources_portion.png} % 替换为你的图片文件名
        \caption{}
        \label{fig:sub1}
    \end{subfigure}
    \hfill
    % 第二张子图 (b)
    \begin{subfigure}[b]{0.4\textwidth}
        \centering
        \includegraphics[width=0.85\textwidth]{figs/gpt_5_mini_retrieved_sources_portion.png} % 替换为你的图片文件名
        \caption{}
        \label{fig:sub2}
    \end{subfigure}
    \hfill
    % 第三张子图 (c)
    \begin{subfigure}[b]{0.4\textwidth}
        \centering
        \includegraphics[width=0.85\textwidth]{figs/deepseek_3_2_retrieved_sources_portion.png} % 替换为你的图片文件名
        \caption{}
        \label{fig:sub3}
    \end{subfigure}
    
    \caption{The distribution of retrieved knowledge sources during \method training.}
    \label{fig:retrieved_training}
\end{figure}
\fi

\begin{figure*}[htbp]
    \centering
    \includegraphics[width=0.98\linewidth]{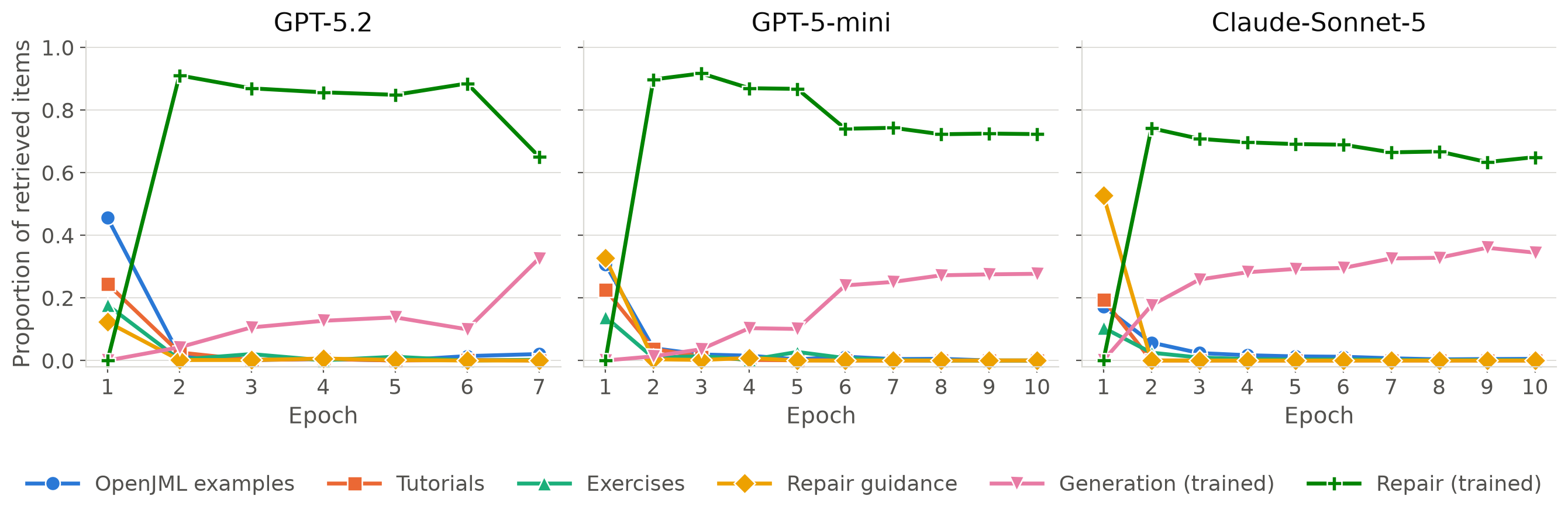}
    \caption{The distribution of retrieved knowledge sources during \method training.}
    \label{fig:retrieved_training}
    \Description{The proportions of six retrieved knowledge sources across training epochs for GPT-5.2, GPT-5-mini, and Claude-Sonnet-5.}
\end{figure*}

\iffalse
\begin{table}[t]
\centering
\begin{tabular}{llccc} % 将 lc 修改为 llc，增加左侧新列
\toprule
    Category & Data Source & GPT-5.2 & GPT-5-mini & DeepSeek-v3.2 \\
\midrule
% 使用 \multirow{行数}{*}{内容}
 \multirow{4}{*}{External} & OpenJML examples & 11 \\
 & OpenJML tutorials & 31 \\
 & OpenJML exercises & 26 \\
 & Repair guidelines \cite{le-cong-etal-2025-llms} & 15 \\
\midrule
 \multirow{2}{*}{Internal} & Generation & 83 \\
 & Repair \\% 将 Total 跨两列对齐
\bottomrule
\end{tabular}
\caption{Distribution of the data sources for the final learned knowledge base of \method. Results are reported in percentage proportions.}
\label{tab:retrieved_inference}
\end{table}
\fi

\begin{table*}[htbp]
\centering
\resizebox{\textwidth}{!}{
\begin{tabular}{lcccccc}
\toprule
    \multirow{2}{*}{Model} & \multicolumn{4}{c}{External} & \multicolumn{2}{c}{Internal} \\
    \cmidrule(lr){2-5} \cmidrule(lr){6-7} % 在主分类下添加细横线区分
    & OpenJML examples & OpenJML tutorials & OpenJML exercises & Repair guidelines \cite{le-cong-etal-2025-llms} & Generation & Repair \\
\midrule
    GPT-5.2       & 1.27 & 0.00 & 0.39 & 0.00 & 34.79 & 63.56\\
    GPT-5-mini    & 0.41 & 0.00 & 0.00 & 0.00 & 28.18 & 71.41\\
    Claude-Sonnet-5 & 0.99 & 0.00 & 0.31 & 0.00 & 37.41 & 61.29 \\
\bottomrule
\end{tabular}
}
\caption{Knowledge type distribution of the number of retrievals during the inference stage. Results are reported in percentage proportions.}
\label{tab:retrieved-inference}
\end{table*}

Figure~\ref{fig:retrieved_training} and Table \ref{tab:retrieved-inference} illustrate the composition of retrieved knowledge over training epochs and inference stage for all three LLM backends. Across all models, the external knowledge sources dominate in the first epoch but rapidly decrease after internally learned knowledge becomes available. Repair-derived items become the predominant source from the second epoch onward. Generation-derived items also become more frequently retrieved in later epochs, especially for GPT-5-mini and Claude-Sonnet-5.
The dominance of learned internal knowledge is more evident in the inference stage. OpenJML official tutorials and expert-written repair guidelines are \textbf{not} retrieved by any of the three LLMs. Internal knowledge accounts for over 98\% of all retrievals on every backend, and repair-derived items are the largest single source. This suggests that during the knowledge training process, the initial external knowledge is gradually ``memorized'' by \method into internal knowledge.
%This shift is consistent across all three backends, though the pace differs. On GPT-5.2 (Figure~\ref{fig:knowledge_distribution}a), learned items begin to dominate around epoch 4--5, whereas on GPT-5-mini (Figure~\ref{fig:knowledge_distribution}b) the transition occurs slightly earlier, likely because the weaker model produces more diverse error patterns that yield a richer set of repair-derived knowledge. On DeepSeek-V3.2 (Figure~\ref{fig:knowledge_distribution}c), a similar trend is observed, with repair-learned items growing particularly fast.

%Notably, among the two types of learned knowledge, repair-derived items consistently account for a larger share than generation-derived items across all models. This suggests that error-fixing trajectories---which encode verifier error messages, erroneous specifications, and their corrections---are more frequently relevant to new specification tasks than successful generation examples alone. This finding aligns with the ablation results in RQ2, where the training mechanism (which produces these learned items) was identified as the most impactful component. Among external sources, tutorials and repair guidances retain a small but persistent presence even in later epochs, indicating that certain foundational JML knowledge remains useful throughout the training process and is not fully supplanted by learned experience.

These results jointly reveal a clear pattern: internally learned knowledge derived from verifier feedback progressively becomes the dominant source of both knowledge storage and retrieval, while external knowledge sources primarily serve as a seed that bootstraps the knowledge learning mechanism in early epochs. This finding further corroborates the ablation results in RQ2, where removing the knowledge evolving mechanism caused one of the largest performance drops.
%claude %, confirming that the true value of \textsc{MemSpec}'s knowledge base lies in its ability to accumulate experiential knowledge from verifier interactions rather than in the static external documents themselves.

%\subsubsection{Analysis on Knowledge Filtering}

\subsubsection{Detailed Analysis on Learned Knowledge Types}

\begin{table*}[t]
\centering
\small
\caption{The detailed taxonomy of knowledge items learned from repair trajectories. Failure types are mutually exclusive; repair strategies are not, as a single repair often performs several actions. Therefore, the repair strategy proportions do not sum to 100\%.}
\label{tab:taxonomy-category}
\resizebox{\textwidth}{!}{%
\begin{tabular}{llp{7.2cm}rrr}
\toprule
\multirow{2}{*}{Taxonomy} & \multirow{2}{*}{Type} & \multirow{2}{*}{Description} & \multicolumn{3}{c}{Proportion} \\
\cmidrule(lr){4-6}
 & & & GPT-5.2 & GPT-5-mini & Claude-Sonnet-5 \\
\midrule
\multirow{10}{*}{Failure type}
 & Arithmetic & Overflow, underflow, range, or other arithmetic proof obligations & 17.9\% & 14.7\% & 22.5\% \\
 & Unsupported Construct & JML constructs not fully supported by OpenJML, such as generalized quantifiers & 28.3\% & 18.3\% & 17.0\% \\
 & Loop Reasoning & Loop invariants or induction-style loop reasoning that fail to verify & 22.1\% & 16.6\% & 24.0\% \\
 & Quantifier Complexity & Solver timeout caused by nested or unbounded first-order quantifiers & 6.7\% & 6.4\% & 8.0\% \\
 & Safety Proof & Null dereference, array index bounds, division by zero, or related safety obligations & 1.7\% & 1.2\% & 4.0\% \\
 & Annotation Structure & Misplaced annotations or incorrectly-placed JML modifiers & 1.2\% & 22.2\% & 3.5\% \\
 & Syntax & Malformed JML syntax, illegal tokens, or ill-formed clauses & 7.5\% & 7.3\% & 7.0\% \\
 & Prover Capacity & Other timeouts that are not caused by a specific construct in the specification & 12.5\% & 11.2\% & 11.0\% \\
 & Verifier Fault & OpenJML internal crashes and unsupported translations outside the specification & 2.1\% & 2.0\% & 3.0\% \\
 & Others & Failures that do not fit any of the above categories & 0.0\% & 0.0\% & 0.0\% \\
\midrule
\multirow{9}{*}{Repair strategy}
 & Remove Unsupported Construct & Remove generalized quantifiers or replace them with verifier-friendly constraints & 40.0\% & 30.6\% & 20.5\% \\
 & Add Numeric Conditions & Add arithmetic assumptions, preconditions, or range constraints & 31.7\% & 48.9\% & 59.5\% \\
 & Add Safety Guards & Add null checks, array bounds, or other runtime-safety conditions & 10.4\% & 21.8\% & 31.0\% \\
 & Rewrite Loop Specification & Rewrite or simplify loop invariants and loop-variant clauses & 67.5\% & 59.2\% & 69.0\% \\
 & Restrict Frame & Add or tighten assignable clauses and method purity annotations & 15.8\% & 19.6\% & 12.5\% \\
 & Weaken Postcondition & Drop or relax postconditions that cannot be proved & 51.2\% & 35.2\% & 33.0\% \\
 & Simplify Quantification & Reduce quantifier nesting or bound a previously unbounded quantifier & 52.1\% & 32.3\% & 28.5\% \\
 & Reformulate Clause Body & Keep the clause structure but rewrite predicates into a prover-friendlier form & 2.5\% & 11.7\% & 6.0\% \\
 & Others & Repairs that do not fit any of the above categories & 0.0\% & 0.2\% & 1.0\% \\
\bottomrule
\end{tabular}
}
\end{table*}

To further understand what kinds of knowledge are learned during \method training, we perform a detailed taxonomy of the most prevalent knowledge type: knowledge built from repair trajectories. We categorize these knowledge items in two repair-related dimensions: 1) why the original specification failed, and 2) how the LLM fixes it. We develop a keyword matching approach to classify the knowledge items by inspecting their content $c_i$ and summary $s_i$ fields. Table~\ref{tab:taxonomy-category} presents a detailed taxonomy of the learned repair knowledge.

\noindent\textbf{Failure patterns.} Across all three LLM backends, three prominent failure types are arithmetic-related errors (14.7--22.5\%), loop reasoning failures (16.6--24.0\%), and the use of constructs not fully supported by OpenJML, such as generalized quantifiers \verb|\num_of| or \verb|\sum| (17.0--28.3\%). Together these three semantic categories account for 49.6\% to 68.3\% of the learned repair knowledge, confirming that the knowledge base concentrates on the reasoning obligations that OpenJML finds hardest rather than on incidental mistakes. The distribution also exposes a clear capability difference between backends. Annotation structure errors, where a JML modifier or a loop specification is rejected because of where it is placed rather than what it says, account for 22.2\% of the knowledge learned from GPT-5-mini but only 1.2\% and 3.5\% on GPT-5.2 and Claude-Sonnet-5. We attribute this to the weaker command of JML surface conventions in lightweight LLMs, which spend a substantial part of their repair budget recovering from annotation placement rather than from failed proofs. Safety obligations and verifier faults each account for less than 5\% across all models, suggesting that once LLMs are augmented with sufficient domain knowledge, low-level mistakes are largely mitigated.

\noindent\textbf{Repair strategies.} The distribution of repair strategies closely mirrors the failure types. Rewriting loop specifications is the most frequent strategy on every backend (59.2--69.0\%), and removing unsupported constructs (20.5--40.0\%) together with adding numeric conditions to guard against arithmetic errors (31.7--59.5\%) directly address the two remaining dominant failure categories. This alignment confirms that the knowledge base captures targeted, actionable fix patterns rather than generic heuristics. Because a single repair usually performs several actions at once, these proportions are not exclusive and do not sum to 100\%; the average repair carries more than two strategy labels. Two strategies reveal how the repairs achieve verifiability. Simplifying quantification (28.5--52.1\%) and weakening postconditions (33.0--51.2\%) both trade specification strength for provability, and they are markedly more frequent on GPT-5.2 (52.1\% and 51.2\%) than on Claude-Sonnet-5 (28.5\% and 33.0\%). Read together with Table~\ref{tab:results}, where Claude-Sonnet-5 attains the higher completeness, this indicates that the strategy mix is not merely a property of the benchmark but reflects how aggressively each backend retreats from a specification it cannot prove. Restricting the frame through \verb|assignable| and purity annotations remains a minority strategy (12.5--19.6\%), consistent with frame conditions being a smaller share of the observed failures.

\begin{tcolorbox}[colback=gray!10,colframe=gray!40,boxrule=0.5pt,
  left=4pt,right=4pt,top=4pt,bottom=4pt]
\noindent\textbf{Answer to RQ3:} Internal knowledge, particularly repair knowledge, dominates the final knowledge base and the retrieval frequency, while external knowledge serves primarily as a seed to empower the knowledge evolving mechanism. The learned repair patterns concentrate on repairing specifications with arithmetic errors, failed loop invariants, and unsupported JML constructs, and the strategies that fix them mirror those same categories. The taxonomy further shows that the composition of learned knowledge is backend-dependent: lightweight LLMs devote a substantially larger share of their repair knowledge to JML annotation structure, whereas stronger backends learn proportionally more about the semantic obligations that make a specification verifiable.
\end{tcolorbox}

\subsection{RQ4: Generalization on SpecGenBench}

Because the distribution of the \method training set FormalBench-Diverse is similar to the evaluation set FormalBench-Base, it is important to investigate whether the knowledge base learned by \method generalizes beyond FormalBench. We further conduct experiments on SpecGenBench, a JML specification generation benchmark introduced by SpecGen~\cite{ma2025specgen}. Same as previous experiments, the knowledge base is evolved only on FormalBench-Diverse and then frozen before evaluation on the 120 SpecGenBench programs. We compare \method with the strongest baseline SpecGen under all three LLM backends, and the results are shown in Table~\ref{tab:specgenbench}.

\begin{table}[H]
\centering
\begin{tabular}{llrrrr}
\toprule
Model & Approach & PR & FR & C@Pass & C@All \\
\midrule
\multirow{2}{*}{GPT-5.2}
 & SpecGen & 76.67 & 18.33 & \textbf{81.20} & 62.26 \\
 & \method & \textbf{90.83} & \textbf{9.17} & 79.65 & \textbf{72.35} \\
\midrule
\multirow{2}{*}{GPT-5-mini}
 & SpecGen & 62.50 & \textbf{3.33} & \textbf{87.57} & 54.73 \\
 & \method & \textbf{88.33} & 11.67 & 81.61 & \textbf{72.09} \\
\midrule
\multirow{2}{*}{Claude-Sonnet-5}
 & SpecGen & 77.50 & \textbf{4.17} & \textbf{80.43} & 62.33 \\
 & \method & \textbf{85.00} & 15.00 & 80.18 & \textbf{68.15} \\
\bottomrule
\end{tabular}
\caption{Results of \method and SpecGen on SpecGenBench. All metrics are reported in \%; C@All is derived from the reported PR and C@Pass.}
\label{tab:specgenbench}
\end{table}

Across all three LLM backends, \method achieves a higher pass rate than SpecGen. The improvements are 14.16, 25.83, and 7.50 percentage points on GPT-5.2, GPT-5-mini, and Claude-Sonnet-5, respectively. On GPT-5.2, \method also reduces the fail rate from 18.33\% to 9.17\%. Although its fail rates are higher on GPT-5-mini and Claude-Sonnet-5, the substantially higher pass rates show that more programs reach successful verification, while the remaining outcomes shift between verifier failures and other statuses such as timeout.

The C@Pass results exhibit the same coverage--completeness tradeoff observed on FormalBench-Base: SpecGen is higher by 1.55, 5.96, and 0.25 percentage points on the three backends. However, \method consistently achieves higher C@All, improving over SpecGen by 10.09, 17.36, and 5.82 percentage points. Therefore, the frozen knowledge base increases the overall yield of complete and verifiable specifications on a benchmark that was not used for knowledge base evolution. This suggests that the learned knowledge captures reusable JML generation and repair patterns rather than being limited to benchmark-specific program instances.

\begin{tcolorbox}[colback=gray!10, colframe=gray!50, boxrule=0.5pt, arc=2pt, boxsep=0pt,]
\textbf{Answer to RQ4:} The knowledge base learned by \method generalizes to SpecGenBench. \method achieves higher PR and C@All than SpecGen across all three LLM backends, indicating that the learned knowledge transfers across specification generation benchmarks.
\end{tcolorbox}

%加个case study?

\section{Threats to Validity}

\textbf{Internal validity.}
A potential threat is the non-determinism in specification generation, as LLMs may introduce randomness. We follow prior work~\cite{le-cong-etal-2025-llms} and use the default value 0.7 for LLMs that support temperature setting, and run 3 repeated experiments. Another threat relates to the design of prompt templates. As the performance of LLMs may be sensitive to prompt design, we use similar templates as previous specification generation approaches \cite{ma2025specgen, le-cong-etal-2025-llms} for specification generation and repair.

\textbf{External validity.}
Our evaluation is conducted exclusively on JML specifications using the OpenJML verifier. While the findings may not directly generalize to other specification languages (e.g., ACSL for C, or Dafny), the design of \method is language-agnostic in principle: the knowledge base initialization and verifier-driven knowledge base evolution can be adapted to any specification language with an automated verifier. Evaluating \method on additional languages and verifiers can be addressed in future work. Another external threat is the scale and complexity of the evaluation datasets. Although the FormalBench dataset only contains standalone Java functions, it is still the largest available JML specification generation benchmark, with complexity higher than previous benchmarks such as SpecGen\cite{ma2025specgen}. In the future, we aim to further extend our specification approach to real-world software projects.

%\textbf{Construct validity.}
%The FormalBench-Diverse dataset, which we use for training, is constructed by applying mutations to programs in FormalBench-Base, which serves as our evaluation set. Although the two sets do not share identical programs, the mutational relationship may introduce a form of distributional similarity that inflates performance. We note that this is the standard experimental setup established by FormalBench~\cite{formalbench}, and all baselines are evaluated under the same conditions. Furthermore, we adopt the same evaluation metrics (pass rate, fail rate, and mutation-based completeness) as prior work, ensuring consistency in how specification quality is measured.

\section{Conclusion}

In this paper, we propose \method, an approach that augments LLMs with a dynamically evolving knowledge base for formal specification generation. \method integrates two complementary knowledge sources: external knowledge from official documentation and tutorials, and internal knowledge distilled from successful generation and repair trajectories via verifier feedback. The knowledge base training mechanism of \method does not require updating LLM parameters, making it agnostic to the choice of LLM and lightweight to deploy. Experiments on the FormalBench benchmark with three LLM backends demonstrate that our approach substantially improves verification pass rates over prior state-of-the-art approaches while preserving specification completeness.

In the future, we plan to extend \method to other formal specification languages and support formal verification for real-world software systems. We also aim to explore more sophisticated knowledge organization strategies, such as hierarchical or graph-structured memory, to improve the efficiency of the training process and the knowledge density of the learned knowledge base.

%\section*{Data Availability}
%Our artifact is available at \url{https://doi.org/10.5281/zenodo.19342584}.

%\begin{acks}

%\end{acks}

\bibliographystyle{ACM-Reference-Format}
\bibliography{sample-base}

%\appendix

\end{document}